\documentclass{article}

\usepackage{amsmath}
\usepackage{graphicx}
\usepackage{subcaption}
\usepackage{setspace}
\usepackage{multirow}
\usepackage{caption}
\usepackage{longtable}
\usepackage{natbib}
\usepackage{amssymb}
\usepackage[title]{appendix}
\usepackage[margin=1in]{geometry}
\usepackage{lineno}
\usepackage{authblk}
\usepackage[nomarkers]{endfloat}
\usepackage[utf8]{inputenc} % usually not needed (loaded by default)
\usepackage[T1]{fontenc}

\title{Anti-diffusive, Non-oscillatory Central (\textit{adNOC}) scheme to
solve the shallow water equations over an erodible substrate in two dimensions.}

\author{
Haseeb Zia
\thanks{Geo-Energy Lab, Gaznat chair on Geo-Energy, École polytechnique fédérale de Lausanne,
EPFL ENAC IIC GEL, 
GC B1 392 (Bâtiment GC), 
Station 18, 
CH-1015 Lausanne, Switzerland.
E-mail: haseeb.zia@epfl.ch}
\thanks{
Dept. of Earth and Environmental Sciences,
Univ. of Geneva,
13 Rue des Maraichers,
1205 Geneva, Switzerland.}

\and Guy Simpson 
\thanks{
Dept. of Earth and Environmental Sciences,
Univ. of Geneva,
13 Rue des Maraichers,
1205 Geneva, Switzerland.
E-mail: Guy.Simpson@unige.ch}
}

\begin{document}
\maketitle
\doublespacing
%\begin{linenumbers}
\begin{abstract}

Shallow water surface flows commonly entrain sediments, resulting in 
scouring and/or deposition of the underlying substrate that may strongly influence the pattern
of subsequent flow. These coupled  phenomena, which can be investigated
mathematically with some extension of the shallow water equations, present
numerous challenges for numerical methods. Here, we present a straightforward
Riemann-solver free approach to solve these equations based on the explicit
non-oscillatory central (NOC) scheme that has already been  widely
applied to hyperbolic conservation laws in other contexts. Our version of 
the central scheme is second-order accurate in time
and space and is used with an anti-diffusive correction to
reduce numerical diffusion usually suffered by central schemes.  Numerical experiments show that
the scheme is accurate and robust for a range of applications from highly dynamic spontaneous
dam break over a mobile bed to slowly evolving morphological bed in an alluvial river.
\end{abstract}

\singlespacing
\doublespacing
\section*{Introduction}
%\label{sec:Introduction}

Sediment transport in surface water flows is a subject with considerable
importance for environmental and engineering problems and is critical
to our understanding and prediction of Earth's surface changes
in response to the gradually evolving climate or to catastrophic extreme
climatic events. The problem is physically complex since it involves
the flow of water that induces erosion or deposition of the underlying substrate,
thereby modifying the surface and subsequent flow
\citep{cao2002mathematical,doi:10.1061,ESP:ESP1157,JFRM:JFRM1098,FLD:FLD853,Simpson:2006:CMS:1650414.1650589,zhiyuan2008,bilanceri:inria-00547550,benkhaldoun2011mathematical,WRCR:WRCR12848,nicholas2013morphodynamic}.
From a mathematical point of view, the problem is also challenging for a number of reasons. First, the system of
governing equations, usually based on some form of the shallow water
equations coupled to sediment transport and bed evolution, is highly
non linear and hyperbolic, having solutions exhibiting propagating
shocks that can cause numerical instabilities with naive schemes
\citep{opac-b1126492,zoppou2003explicit}. Second, as the water depth approaches zero close to
the wet/dry front, it can cause negative water depth due to small numerical oscillations, resulting
in break down of calculations \citep{audusse2005well,kurganov2007second,beljadid2016well}. Diminishing water height can also cause
the equations to become singular (due to bed
friction), which can lead to numerical instabilities \citep{bradford2002finite,begnudelli5276784,Bollermann2012}.
 Third, accurately calculating the seemingly trivial steady state of water body at rest over
 irregular bottom topography have proved to be challenging. In this case, a delicate balance is required between the 
 flux gradient and the source term in the momentum conservation equation. Even a small imbalance
 between the two terms results in unphysical spontaneous water movement (the so called
 numerical storm). Schemes that are able to accurately calculate steady state solution over
 irregular topography are referred as C-property maintaining or well-balanced schemes (some
 examples of such schemes are
 \citep{leveque1998balancing,rogers2003mathematical,castro2005numerical,Canestrelli2010291,kesserwani2010well,ricchiuto2011c,Bollermann2012,bollermann2015finite}).
 Fourth, if the interaction between the flow of water and evolution of the mobile bed is weak, the
 characteristic time scales of the flow and of the sediment transport can be very different,
 causing the equations to become stiff (e.g., see \citep{bilanceri:inria-00547550}),
 necessitating very small time steps to ensure stability with explicit methods.

In the last few decades, numerous high resolution numerical methods have been
developed and applied to solve systems of hyperbolic conservation laws such as the shallow water system (e.g.,
\citep{Harten1983357,sweby1984,opac-b1126492,leveque2002finite,Jiang98nonoscillatorycentral}).
Many of these schemes employ the Godunov approach by which the approximate
solution is realised by a piecewise polynomial that is reconstructed from the evolving cell-averages. Within the Godunov class of methods, two main approaches can be distinguished: upwind and centered. Many upwind methods evaluate
numerical fluxes across cell boundaries using knowledge of the wave
structure (given by the eigen values of the Jacobian of the system) in combination with Riemann
solvers (e.g., \citep{doi:10.1061,FLD:FLD853,Simpson:2006:CMS:1650414.1650589,zhiyuan2008,bilanceri:inria-00547550,benkhaldoun2011mathematical,WRCR:WRCR12848,Nicholas2013,gmdd-7-2429-2014,FLM:111083,WRCR:WRCR12848})
or ENO or WENO reconstruction (e.g.,
\citep{Crnjaric-Zic:2004:BFV:1044287.1044293,MR2383221}). These methods are
often highly accurate, but can suffer from splitting and are less successful in cases where the detailed wave structure of the equations is poorly known (as can be the case for the type of problem investigated here).
Centered schemes on the other hand - of which CWENO and NOC are two
examples - tend to be simpler since they normally require only the largest wave speed while they
can also have high resolution comparable with upwind schemes \citep{Caleffi2007730,Canestrelli2010291}.

In this paper we present an anti-diffusive, non-oscillatory central 
(\textit{adNOC}) scheme to solve the 2D shallow water equations coupled to an erodible
substrate that is straightforward, robust and accurate. The scheme is based on the well
balanced scheme presented in \citep{zia2016} which is an extension of the non
oscillatory central (NOC) scheme (see
\citep{NessyahuTadmor1990,Jiang98nonoscillatorycentral,doi:10.1137/S0036142997317560}), which
itself is a higher order extension of the classical first order centered
Lax-Friedrichs scheme. The extended NOC scheme, referred as \textit{adNOC} uses an
anti-diffusive correction to reduce numerical dissipation suffered by central schemes when solving
coupled systems (see \citep{zia2015numerical} for detail). The numerical dissipation arises due to
difference in characteristic time scales of the coupled systems which causes the relatively less
dynamic processes to be solved with small time steps, for which NOC schemes are known to be
diffusive
\citep{kurganov2000new,huynh2003analysis,kurganov2007reduction,siviglia2013numerical,canestrelli2012restoration,stecca2012finite}.
Although the method presented is second-order accurate in time and space, higher order cell
reconstructions could be employed, if desired, to increase accuracy (e.g.,
\citep{Harten1983357,ZAMM:ZAMM200310123}).
The \textit{adNOC} scheme presented here has additional
advantages over other Riemann-solver free central schemes such as the central-upwind schemes
\citep{kurganov2002central,kurganov2007second,bryson2011well,Bollermann2012,bollermann2013well,liu2015well,beljadid2016well}
and \textit{PRICE} schemes \citep{canestrelli2009well,Canestrelli2010291} as it 
requires the wave structure of the system only to estimate the upper bound on the time step and not
to calculate fluxes. For the coupled system being solved here, the upper bound calculated with
the largest wave speed of the shallow water equations is found to be sufficient for stability
in almost all cases (\cite{cordier2011bedload}). Hence the system is solved withoug the Jacobian and
eigen values of the sediment transport component. This provides the much needed flexibility to
change the empirical relations  
to calculate sediment fluxes on case to case basis,
without having to calculate their Jacobian.

 In what follows, we present the governing equations, details of the numerical scheme (based largely on the
formulation presented by \citep{pudasaini2007avalanche} for modelling granular
flows) and anti-diffusion correction, along with
several computed test cases.

\section*{Governing Equations}
%\label{sec:Governing Equations}

The equations governing 2D shallow water flow coupled to an erodible
substrate comprise of mass and momentum conservation equations for the
water-sediment mixture and the mass conservation equations for the sediment and bed material. The equations are written as:

\begin{equation}
\frac{\partial h}{\partial t}+\frac{\partial(hu)}{\partial x}+\frac{\partial(hv)}{\partial
y}=-\frac{\partial z}{\partial t},\label{eq:1}
\end{equation}

\begin{equation}
\frac{\partial(hu)}{\partial t}+\frac{\partial}{\partial
x}\left(hu^{2}+\frac{1}{2}gh^2\right)+\frac{\partial}{\partial
y}(huv)=B_{x},\label{eq:2}
\end{equation}

\begin{equation}
\frac{\partial(hv)}{\partial t}+\frac{\partial}{\partial x}(huv)+\frac{\partial}{\partial
y}\left(hv^{2}+\frac{1}{2}gh^{2}\right)=B_{y},\label{eq:3}
\end{equation}

\begin{equation}
\frac{\partial(hc)}{\partial t}+\frac{\partial(hcu)}{\partial
x}+\frac{\partial(hcv)}{\partial y}=E-D,\label{eq:4}
\end{equation}

\begin{equation}
\frac{\partial z}{\partial t}+\frac{1}{1-\phi}\frac{\partial
q_{x}}{\partial x}+\frac{1}{1-\phi}\frac{\partial
q_{y}}{\partial y}=\frac{D-E}{1-\phi},\label{eq:5}
\end{equation}

\noindent where $B_{x}$ and $B{}_{y}$ are source/sink terms defined as

\begin{equation}
B_{x}=-gh\frac{\partial z}{\partial
x}-ghS_{fx}-\frac{(\rho_{s}-\rho_{w})gh^{2}}{2\rho}\frac{\partial c}{\partial
x}+\frac{(\rho_{0}-\rho)u}{\rho}\frac{\partial z}{\partial
t},\label{eq:6}
\end{equation}

\begin{equation}
B_{y}=-gh\frac{\partial z}{\partial
y}-ghS_{fy}-\frac{(\rho_{s}-\rho_{w})gh^{2}}{2\rho}\frac{\partial c}{\partial
y}+\frac{(\rho_{0}-\rho)v}{\rho}\frac{\partial z}{\partial
t}.\label{eq:7}
\end{equation}
\doublespacing
Similar equations have
been presented by
\citep{Fagherazzi20031143,doi:10.1061,Simpson:2006:CMS:1650414.1650589,xia2010modelling,yue2008two}.
In these equations, $t$ is the time, $x$ and $y$ are horizontal
coordinates, $h$ is the flow depth, $u$ and $v$ are depth-averaged
velocities in the $x$ and $y$ directions respectively, $z$ is the
bed elevation, $c$ is the depth-averaged volumetric sediment concentration,
$g$ is the gravitational acceleration, $S_{fx}$ and $S_{fy}$ are friction
slopes in the $x$ and $y$ directions respectively, $\phi$ is the bed
sediment porosity, $E$ and $D$ are substrate entrainment and deposition
fluxes across the bottom boundary of flow (representing sediment exchange
between the water column and bed), $\rho=\rho_{w}(1-c)+\rho_{s}c$ is the density
of the water-sediment mixture, $\rho_{0}=\rho_{w}\phi+\rho_{s}(1-\phi)$ is the
density of the saturated bed, $\rho{}_{s}$, $\rho{}_{w}$ are the densities of water and sediment
respectively and $q_x$, $q_y$  are bed load fluxes in the x and y directions respectively (see table
\ref{table:Notations} for a summary of  notation).

Eq. (\ref{eq:1}) and Eq. (\ref{eq:4}) represents the mass conservation for water
and sediment whereas Eq. (\ref{eq:2}) and Eq. (\ref{eq:3}) represents momentum
conservation in $x$ and $y$ directions respectively. The terms
on the left side of Eq. (\ref{eq:2}) and (\ref{eq:3})  account for inertia and
pressure effects in the flowing fluid. The terms $B_{x}$ and $B_{y}$ are the
source terms which are expanded in Eq. (\ref{eq:6}) and (\ref{eq:7}) respectively.
In Eq. (\ref{eq:6}) and (\ref{eq:7}), the terms from left to right, account for bed topography,
friction loss, spatial variations in sediment concentration and momentum transfer between
the flow and the erodible bed. The first two terms are part of the classical
clear-water equations whereas the last two terms only become important in
highly concentrated flows and can be neglected in situations where the sediment
concentration is low. Dispersive sediment transport has been neglected in the
model but can be included by adding additional diffusion terms. The sediment mass
conservation represented by Eq. (\ref{eq:4}) signifies the suspended component
of sediment being transported, which increases
when the local erosive flux $E$ exceeds the depositional flux $D$. Eq. (\ref{eq:5}) is the Exner equation (e.g.
see \citep{paola2005generalized}) which accounts for the change in the bed
elevation as a result of variations in bed load fluxes $q_x$, $q_y$ and the local sediment
erosive and depositional fluxes.

In order to close the governing equations, it is necessary to specify
additional relations for the friction slope, the substrate exchange
fluxes between water and bed, and the bedload fluxes. For the friction slope, several
classical equations exist, suitability of which depends on the
flow conditions. In this study we have used Manning's equation to approximate
friction loss for turbulent flows:

\begin{equation}
S_{fx}=\frac{n^{2}u\sqrt{u^{2}+v^{2}}}{h^{4/3}},\;
S_{fy}=\frac{n^{2}v\sqrt{u^{2}+v^{2}}}{h^{4/3}},\label{eq:8}
\end{equation}
where $n$ is the Manning\textquoteright{}s roughness coefficient. For quantifying
entrainment and deposition fluxes, a large number of empirical relations have been proposed, a
review of which can be found in \citep{cao2002mathematical}. For
deposition of sediment, we are using the following relation:

\begin{equation}
D=\omega(1-C_{a})^{i}C_{a},\label{eq:9}
\end{equation}

\noindent where $\omega$ is the settling velocity of a single particle in tranquil
water given by:

\[
\omega=\sqrt{(13.95\nu/d)^{2}+1.09\rho_{s}gd}-13.95\nu/d,
\]
$\nu$ is the kinematic viscosity of water, $d$ is the grain diameter,
$g$ is the gravitational acceleration, $\rho_{s}$ is the sediment
particle density, $C_{a}$ is the near-bed volumetric sediment concentration
and $i$ is an exponent (2 in this study). The value for $C_{a}$ is computed from
the relation $C_{a}=\alpha_{c}c$ where $c$ is the depth-averaged volumetric
sediment concentration and $\alpha_{c}$ is a coefficient larger than
unity. In order so that the near-bed concentration does not exceed
$(1-\phi)$(where $\phi$ is the bed sediment porosity), the coefficient
$\alpha_{c}$ is computed using $\alpha_{c}=min(2,(1-\phi)/c)$(see
\citep{doi:10.1061}).

For calculation of erosion, we have used the following relation
(see
\citep{doi:10.1061}):

\begin{equation}
E=\zeta\frac{160}{R^{0.8}}\frac{(1-\phi)}{\theta_{c}}\frac{d(\theta-\theta_{c})U_{\infty}}{h},\label{eq:10}
\end{equation}

\noindent where $R=d\sqrt{sgd}/\nu$, $d$ is the
sediment grain size, $s=\rho_{s}/\rho_{w}-1$, $\nu$ is the kinematic viscosity of water, $\theta$ is
Shields parameter $=u_{\star}/(sgd)$, $u_{\star}$ is the friction
velocity $(=\sqrt{f/8}\sqrt{u^{2}+v^{2}})$, $f$ is Darcy-Weisbach
friction factor, $\theta_{c}$ is the critical value of the
Shields parameter for the initiation of sediment motion below which
$E=0$, $U_{\infty}$ is the free surface velocity $=7\sqrt{u^{2}+v^{2}}/6$.
Since the erosion formulation proposed in \citep{doi:10.1061} approximates the total sediment flux
including both suspended load and bedload, we have added a factor $\zeta$ in Eq.
(\ref{eq:10}) bounded by $0\leq\zeta\leq 1$ signifying the suspended
load component of the total sediment transport. This is done due to the fact that the model has 
explicit bedload flux terms in the Exner equation.

Same as the erosion and deposition, there is a large
number of empirical formulations present in literature to approximate the bedload flux (e.g. see
\citep{yang2006erosion} for a detailed review).
%These formulations are mostly based on shear stress exerted on the sediment by the flow (see \citep{yang2006erosion} for a detailed review).
 In this
study we use the shear stress free formulation discussed by Grass \citep{grass1981}
given by:
\begin{equation}
q_{x}=Au(\sqrt{u^{2}+v^{2}})^{b-1},\;
q_{y}=Av(\sqrt{u^{2}+v^{2}})^{b-1},\label{eq:11}
\end{equation}
where \textit{A} is a constant 
signifying the interaction of the sediment with the flow and  $b$ is
an exponent ($1\leq b \leq 4$). Although being fairly basic, as it does not
assume any critical threshold velocity to initiate bedload transport,
the Grass formulation has been used by many previous
studies (e.g.
\citep{liu2008two,benkhaldoun2010two,vcrnjaric2004extension,siviglia2013numerical}).

There are several advantages of the sediment transport formulation
such as that presented here (see also
\citep{FLM:14463,doi:10.1061,Simpson:2006:CMS:1650414.1650589}). First, the
evaluation of entrainment and deposition is done independently of each other.
This is not only convenient but it makes good sense given that erosion and
sedimentation are governed by completely different physics.
In contrast to the sediment capacity approach where the change in local bed
morphology is evaluated by considering whether the sediment discharge is greater or less than
the sediment transport capacity, the change in bed morphology is evaluated
by taking the difference between the local erosion and deposition fluxes.
Thus, there is no need for assumptions concerning whether sediment transport is
supply-limited or transport-limited. Second, the
distinction between the bedload and suspended load is made, which is necessary because the
suspended load and bedload are governed by different physics. For example,
bedload transport is affected by bottom slope while suspended sediment load is
not. Moreover, the suspended sediment is transported with the velocity of the
flow while bedload is usually transported at a slower rate. Finally, because an
attempt is made to clearly separate the governing equations from the empirical relations, the
formulae for calculation of the substrate exchange and bedload fluxes can readily be modified
to incorporate new transport models or to study different applications. The advantage of such
formulation is fully availed by the central scheme presented here. For the upwind schemes, in
comparison, a change in the empirical formulation changing the
wavestructure of the system will require analytical relations for the new
wave speeds or their numerical calculation which is computationally costly.

\section*{NOC Numerical Scheme}
%\label{sec:NOC Numerical Scheme}

The numerical scheme utilized to solve the equations is based on the non-oscillatory
central (NOC) scheme developed by
\citep{NessyahuTadmor1990} (see also
\citep{doi:10.1137/S0036142997317560,Jiang98nonoscillatorycentral}).
 The scheme uses a two step predictor-corrector procedure. The first step
 involves first order prediction of the grid values according to non-oscillatory reconstructions
 from given cell averages while the second step involves staggered averaging to determine the full evolution
 of these averages. Below, we present details, based largely on the formulation presented by
 \citep{pudasaini2007avalanche}, of how this scheme can be used to solve the two dimensional shallow
 water equations coupled to a mobile substrate.

For the purposes of obtaining a numerical solution we write the governing
equations in the vector form as follows:

\begin{equation}
\frac{\partial W}{\partial t}+\frac{\partial F}{\partial x}+\frac{\partial
G}{\partial y}=S,\label{eq:12}
\end{equation}

\noindent where $W$ is the solution vector defined as:

\begin{equation}
W=\left[\begin{array}{c}
h\\
hu\\
hv\\
hc\\
z
\end{array}\right],\label{eq:13}
\end{equation}

\noindent $F$ and $G$ are flux vectors defined as:

\begin{equation}
F=\left[\begin{array}{c}
hu\\
hu^{2}+\frac{1}{2}gh^{2}\\
huv\\
huc\\
\frac{1}{1-\phi}q_x
\end{array}\right],\label{eq:14}
\end{equation}

\begin{equation}
G=\left[\begin{array}{c}
hv\\
huv\\
hv^{2}+\frac{1}{2}gh^{2}\\
hvc\\
\frac{1}{1-\phi}q_y
\end{array}\right],\label{eq:15}
\end{equation}

and $S$ is the source vector defined by:

\begin{equation}
S=\left[\begin{array}{c}
-\frac{\partial z}{\partial t}\\
B_x\\
B_y\\
E-D\\
\frac{D-E}{1-\phi}
\end{array}\right].\label{eq:16}
\end{equation}

\begin{figure}
\centering
\includegraphics[scale=0.4]{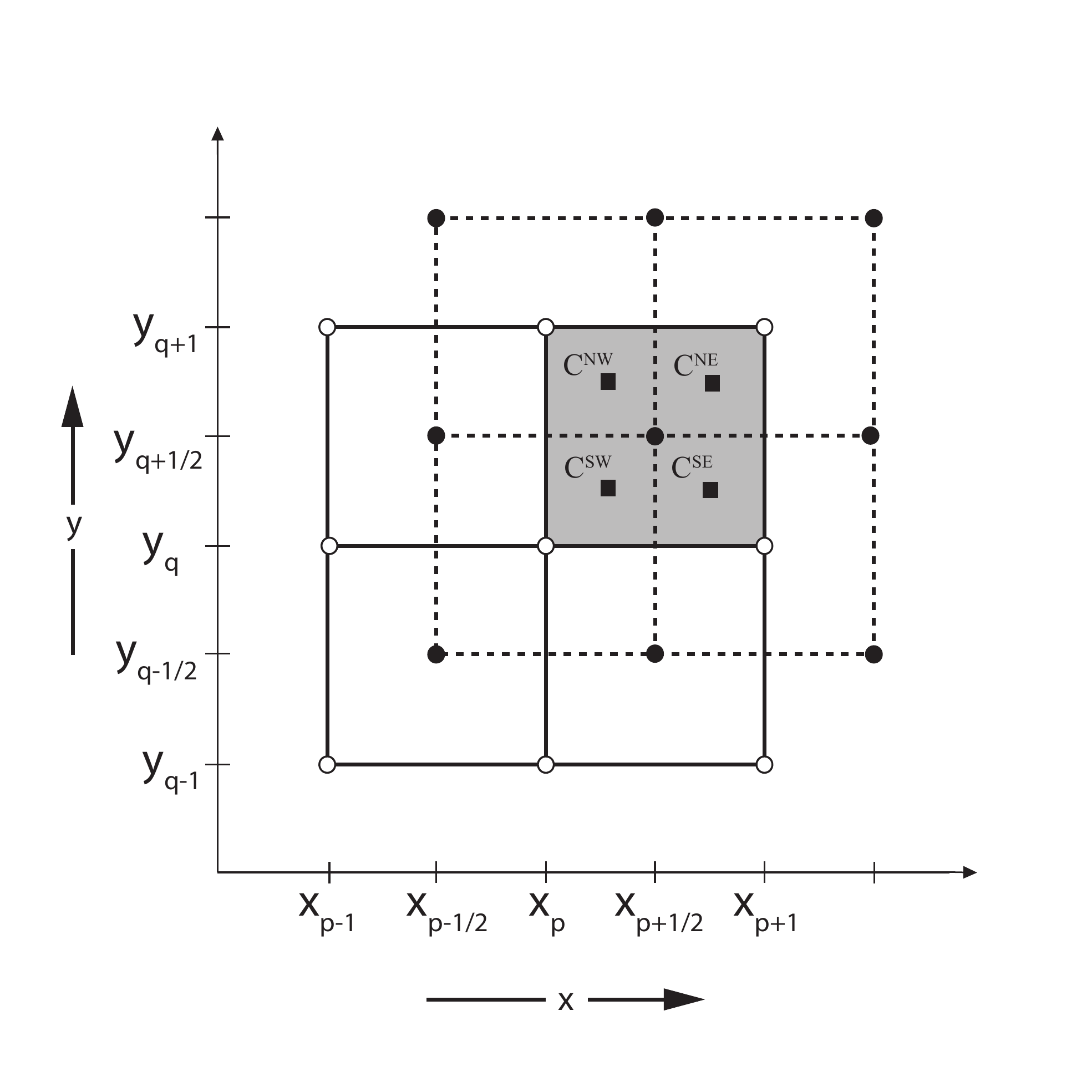}

\caption{Staggered grid with grid points marked by $\circ$ and $\bullet$ for
the two different grids. The cell centers of the contributing sub-cells are
marked by $^\blacksquare$.}
\label{fig:grid}
\end{figure}

We begin by dividing  the two dimensional spatial domain  into rectangular cells (see Fig. 1).  Let
$C_{p,q}$ denote the cell that covers the region $ |x-x_{p}|\le\frac{\text{\ensuremath{\Delta}}x}{2},|y-y_{q}|\le\frac{\Delta
y}{2},$ where $p$ and $q$ are integer node indices and $\Delta x$ and $\Delta y$
are grid spacings in the $x$ and $y$ directions respectively. For the
development below, we note that  the  cell  $C_{p+1/2,q+1/2}$ consists of the
overlap between four intersecting cells $C_{p,q}$, $C_{p+1,q}$, $C_{p+1,q+1}$,
$C_{p,q+1}$, denoted  $C^{SW}$,  $C^{SE}$,  $C^{NE}$, 
$C^{NW}$ (see Figure \ref{fig:grid}). Let $\overline{W}{}_{p,q}^{n}$ denote the
cell average over the cell at time $t^{n}$. The solution can be reconstructed in space linearly over the cell from the
average  by: 

\begin{equation}\label{eq:17}
W_{p,q}(x,y,t^{n})=\overline{W}{}_{p,q}^{n}+\sigma_{p,q}^{x}(x-x_{p})+\sigma_{p,q}^{y}(y-y_{q}),\,\hspace{15pt}
(x,y) \; \epsilon \; C_{p,q},
\end{equation}
\noindent where $\sigma^{x}$ and $\sigma{}^{y}$ are the discrete slopes of
$W$ in the $x$ and $y$ directions, respectively. This reconstruction
is used to achieve second-order accuracy in space. Second-order
temporal accuracy is achieved by using a predictor-corrector procedure in which the solution is first evaluated at the half time step in the predictor step. Linear reconstruction is also used for reconstruction in time:
\[
\overline{W}_{p,q}^{n+1/2}=\overline{W}_{p,q}^{n}+\frac{\Delta
t}{2}\left(\frac{\partial W}{\partial t}\right)^{n},
\]
where $\left(\frac{\partial W}{\partial t}\right)^{n}$ is calculated using the
conservation law (i.e., Eq. \ref{eq:12}):

\begin{equation}\label{eq:18}
\left(\frac{\partial W}{\partial t}\right)^{n}=-\left(\frac{\partial
F(W)}{\partial x}\right)^{n}-\left(\frac{\partial G(W)}{\partial
y}\right)^{n}+S(\overline{W}^{n}).
\end{equation}
Thus the predictor step is given by:
\begin{equation}\label{eq:19}
\overline{W}_{p,q}^{n+1/2}=\overline{W}_{p,q}^{n}-\frac{\Delta
t}{2}(\sigma^{F})_{p,q}^{n}-\frac{\Delta
t}{2}(\sigma^{G})_{p,q}^{n}+\frac{\Delta t}{2}S(\overline{W}_{p,q}^{n}),
\end{equation}
where $\sigma^{F}$ and $\sigma^{G}$ are one-dimensional discrete slopes of the
fluxes $F$ and $G$ in the $x$ and $y$ directions, respectively.

To calculate the second order solution for the conservation law given by Eq.
(\ref{eq:12}),  we integrate it over the cell $C_{p+\frac{1}{2},q+\frac{1}{2}}$
and time period {[}$t^{n},t^{n+1}${]}. The staggered average over
$C_{p+\frac{1}{2},q+\frac{1}{2}}$ is given by:

\[
\overline{W}_{p+1/2,q+1/2}^{n+1}=\frac{1}{\Delta x\Delta
y}\int_{x_{p}}^{x_{p+1}}\int_{y_{q}}^{y_{q+1}}W(x,y,t^{n})dxdy
-\frac{1}{\Delta x\Delta y}\int_{t^{n}}^{t^{n+1}}\int_{y_{q}}^{y_{q+1}}(F(x_{p+1},y,t)-F(x_{p},y,t))dydt
\]

\begin{equation}\label{eq:20}
-\frac{1}{\Delta x\Delta y}\int_{t^{n}}^{t^{n+1}}\int_{x_{p}}^{x_{p+1}}(G(x,y_{q+1},t)-G(x,y_{q},t))dxdt
+\frac{1}{\Delta x\Delta
y}\int_{t^{n}}^{t^{n+1}}\int_{x_{p}}^{x_{p+1}}\int_{y_{q}}^{y_{q+1}}S(x,y,t)dxdydt.
\end{equation}

We now consider how to separately compute each of the terms in this
equation (\ref{eq:20}). The first term on the right hand side can be
split into four parts representing the individual contributions from the four adjacent
intersecting cells on the staggered grid (see Fig. \ref{fig:grid}):

\[
\int_{x_{p}}^{x_{p+1}}\int_{y_{q}}^{y_{q+1}}W(x,y,t^{n})dxdy=
\int\int_{C^{SW}}W(x,y,t^{n})dxdy+\int\int_{C^{SE}}W(x,y,t^{n})dxdy
\]

\begin{equation}\label{eq:21}
+\int\int_{C^{NE}}W(x,y,t^{n})dxdy+\int\int_{C^{NW}}W(x,y,t^{n})dxdy.
\end{equation}

\noindent To evaluate this integral, we begin by integrating the reconstructing polynomial
(Eq. \ref{eq:17}) over the sub-cell $C^{SW}$. The integral is given by:

\[
\int\int_{C^{SW}}W(x,y,t^{n})=\int_{x_{p}}^{x_{p+1/2}}\int_{y_{q}}^{y_{q+1/2}}(\overline{W}_{p,q}^{n}+\sigma_{p,q}^{x}(x-x_{p})+\sigma_{p,q}^{y}(y-y_{q}))dxdy.
\]

\noindent Calculating the definite integral and dividing by $\Delta x\Delta y$
to get the cell average:

\[
\frac{1}{\Delta x\Delta
y}\int\int_{C^{SW}}W(x,y,t^{n})=\frac{1}{4}\overline{W}_{p,q}^{n}+\frac{\Delta
x}{16}\sigma_{p,q}^{x}+\frac{\Delta y}{16}\sigma_{p,q}^{y}.
\]

\noindent The other terms in Eq. (\ref{eq:21})  can be calculated by the same
procedure, leading to:

\[
\frac{1}{\Delta x\Delta
y}\int\int_{C^{NW}}W(x,y,t^{n})=\frac{1}{4}\overline{W}_{p,q+1}^{n}+\frac{\Delta
x}{16}\sigma_{p,q+1}^{x}-\frac{\Delta y}{16}\sigma_{p,q+1}^{y},
\]

\[
\frac{1}{\Delta x\Delta
y}\int\int_{C^{NE}}W(x,y,t^{n})=\frac{1}{4}\overline{W}_{p+1,q+1}^{n}-\frac{\Delta
x}{16}\sigma_{p+1,q+1}^{x}-\frac{\Delta y}{16}\sigma_{p+1,q+1}^{y},
\]

\[
\frac{1}{\Delta x\Delta
y}\int\int_{C^{SE}}W(x,y,t^{n})=\frac{1}{4}\overline{W}_{p+1,q}^{n}-\frac{\Delta
x}{16}\sigma_{p+1,q}^{x}+\frac{\Delta y}{16}\sigma_{p+1,q}^{y}.
\]

\noindent Next, we compute the second and third terms in Eq. (\ref{eq:20}), i.e. the
integral of fluxes through the four edges of each
cell. The fluxes at the northern and southern edges are
associated with the flux function $G$ whereas the fluxes at the eastern and western edges are
associated with the flux function $F$. To calculate the integral, we use the
midpoint quadrature  rule. For example, the flux integral at the western edge is given by:

\[
\frac{1}{\Delta x\Delta y}\int_{t^{n}}^{t^{n+1}}\int_{y^{q}}^{y^{q+1}}F(x_{p},y,t)dydt=\frac{\Delta t}{2\Delta x}(F(x_{p},y_{q},t^{n+1/2})+F(x_{p},y_{q+1},t^{n+1/2}))
\]

\[
=\frac{\Delta t}{2\Delta
x}(F(\overline{W}_{p,q}^{n+1/2})+F(\overline{W}_{p,q+1}^{n+1/2})).
\]

\noindent Similarly, the integral of fluxes respectively through the eastern, southern and northern edges
of the cell are:

\[
\frac{1}{\Delta x\Delta
y}\int_{t^{n}}^{t^{n+1}}\int_{y^{q}}^{y^{q+1}}F(x_{p+1},y,t)dydt=\frac{\Delta t}{2\Delta x}(F(\overline{W}_{p+1,q}^{n+1/2})+F(\overline{W}_{p+1,q+1}^{n+1/2})),
\]

\[
\frac{1}{\Delta x\Delta
y}\int_{t^{n}}^{t^{n+1}}\int_{x^{p}}^{x^{p+1}}G(x,y_{q},t)dxdt=\frac{\Delta t}{2\Delta y}(G(\overline{W}_{p,q}^{n+1/2})+G(\overline{W}_{p+1,q}^{n+1/2})),
\]

\[
\frac{1}{\Delta x\Delta
y}\int_{t^{n}}^{t^{n+1}}\int_{x^{p}}^{x^{p+1}}G(x,y_{q+1},t)dxdt=\frac{\Delta t}{2\Delta y}(G(\overline{W}_{p,q+1}^{n+1/2})+G(\overline{W}_{p+1,q+1}^{n+1/2})).
\]

\noindent Note that the solution at half time step is evaluated in
the predictor step (Eq. \ref{eq:19}). Finally,  we compute the last remaining
term in Eq. (\ref{eq:20}) i.e. the integration of the source term. The integral is computed by adding
up contributions from the four neighbouring cells i.e. the sub-cells $SC^{SW}, SC^{SE}, SC^{NE}$ and
$SC^{NW}$. The source is evaluated at the centers
of the contributing sub-cells and the average of the four is taken for
calculation of the predictor step:

\[
\frac{1}{\Delta x\Delta
y}\int_{t^{n}}^{t^{n+1}}\int_{x_{p}}^{x_{p+1}}\int_{y_{q}}^{y_{q+1}}S(x,y,t)dxdydt=\frac{\Delta
t}{4}\{S(\overline{W}_{p+1/4,q+1/4}^{n+1/2})+S(\overline{W}_{p+3/4,q+1/4}^{n+1/2})
\]

\[
+S(\overline{W}_{p+1/4,q+3/4}^{n+1/2})+S(\overline{W}_{p+3/4,q+3/4}^{n+1/2})\}.
\]
The value of the solution average $\overline{W}^{n+1/2}$ at the sub-cells
centers (represented by $^\blacksquare$ in Fig. \ref{fig:grid}) are computed by
utilizing Taylor series expansions:

\[
\overline{W}_{p+1/4,q+1/4}^{n+1/2}=\overline{W}_{p,q}^{n+1/2}+\frac{\Delta
x}{4}(\sigma^{x})_{p,q}^{n}+\frac{\Delta y}{4}(\sigma^{y})_{p,q}^{n},
\]

\[
\overline{W}_{p+3/4,q+1/4}^{n+1/2}=\overline{W}_{p+1,q}^{n+1/2}-\frac{\Delta
x}{4}(\sigma^{x})_{p+1,q}^{n}+\frac{\Delta y}{4}(\sigma^{y})_{p+1,q}^{n},
\]

\[
\overline{W}_{p+3/4,q+3/4}^{n+1/2}=\overline{W}_{p+1,q+1}^{n+1/2}-\frac{\Delta
x}{4}(\sigma^{x})_{p+1,q+1}^{n}-\frac{\Delta y}{4}(\sigma^{y})_{p+1,q+1}^{n},
\]

\[
\overline{W}_{p+1/4,q+3/4}^{n+1/2}=\overline{W}_{p,q+1}^{n+1/2}+\frac{\Delta
x}{4}(\sigma^{x})_{p,q+1}^{n}-\frac{\Delta y}{4}(\sigma^{y})_{p,q+1}^{n}.
\]

\noindent Collecting all the terms in the Eq. (\ref{eq:20}) results in the standard second-order,
non-oscillatory central scheme given by:

\[
\overline{W}_{p+1/2,q+1/2}^{n+1}=\frac{1}{4}\{\overline{W}_{p,q}^{n}+\overline{W}_{p+1,q}^{n}+\overline{W}_{p,q+1}^{n}+\overline{W}_{p+1,q+1}^{n}\}
\]

\[
+\frac{\Delta x}{16}\{\sigma_{p,q}^{x}-\sigma_{p+1,q}^{x}-\sigma_{p+1,q+1}^{x}+\sigma_{p,q+1}^{x}\}
\]

\[
+\frac{\Delta y}{16}\{\sigma_{p,q}^{y}+\sigma_{p+1,q}^{y}-\sigma_{p+1,q+1}^{y}-\sigma_{p,q+1}^{y}\}
\]

\[
-\frac{\Delta t}{2\Delta x}\{F(\overline{W}_{p+1,q}^{n+1/2})+F(\overline{W}_{p+1,q+1}^{n+1/2})-F(\overline{W}_{p,q}^{n+1/2})-F(\overline{W}_{p,q+1}^{n+1/2})\}
\]

\[
-\frac{\Delta t}{2\Delta y}\{G(\overline{W}_{p,q+1}^{n+1/2})+G(\overline{W}_{p+1,q+1}^{n+1/2})-G(\overline{W}_{p,q}^{n+1/2})-G(\overline{W}_{p+1,q}^{n+1/2})\}
\]

\begin{equation}\label{eq:22}
+\frac{\Delta
t}{4}\{S(\overline{W}_{p+1/4,q+1/4}^{n+1/2})+S(\overline{W}_{p+3/4,q+1/4}^{n+1/2})+S(\overline{W}_{p+3/4,q+3/4}^{n+1/2})+S(\overline{W}_{p+1/4,q+3/4}^{n+1/2})\}.
\end{equation}

It is important to remember that while the high order nature of the scheme assures
that shocks and discontinuities are captured, this comes at the expense of
spurious oscillations. To avoid these oscillations, NOC scheme uses the well-known Total
Variation Diminishing (TVD) concept (see \citep{Harten1983357}) by calculating the slopes
$\sigma{}^{x}$, $\sigma{}^{y}$ and the flux slopes $\sigma^{F}$, $\sigma^{G}$
using limiters (e.g.
\textit{min-mod},  \textit{superbee}, etc.). In this study, we have used the \textit{min-mod}
limiter given by:
\[
minmod\{r_1,r_2\}=\frac{1}{2}[sgn(r_1)+sgn(r_2)].Min(|r_1|,|r_2|),
\]
where $r_1$ and $r_2$ are the slopes at successive positions on the solution
mesh in any given direction.

Because the numerical scheme presented is entirely
explicit, the maximum allowable time step for stability is constrained by the
Courant-Friedrichs-Lewy condition. In this study we compute the time step
dynamically according to the condition:
\[
\Delta t=min(\Delta x \; Cn /V_{max}, \Delta y \; Cn /V_{max}),
\]
where $Cn$ is the Courant number and $ V_{max} $ is the maximum wave speed. The
Courant number must have a value of unity or less in order for the scheme to be numerically stable.
For the maximum wave speed, we are using the largest eigen value of the shallow
water equations:
\[
V_{max}=max(|u-\sqrt{gh}|,|u+\sqrt{gh}|,|v-\sqrt{gh}|,|v+\sqrt{gh}|).
\]
As noted by \cite{cordier2011bedload}, estimating maximum wave speed of the coupled shallow
water/sediment transport system with the largest eigen value of the shallow water equation can be
problematic for splitting schemes, but is sufficient for Lax-Friedrich scheme based central schemes.
Hence for the \textit{adNOC} scheme, the solution is evaluated without any knowledge of the wave
structure of the sediment transport system.

\section*{Anti-diffusion correction}
%\label{sec:Anti-diffusion correction}

Despite the advantages of simplicity and universality,
there are certain cases where use of the standard central scheme described above becomes unfeasible.
Central schemes are known to be excessively diffusive, especially when small time steps are used
\citep{kurganov2000new,huynh2003analysis,kurganov2007reduction,siviglia2013numerical,canestrelli2012restoration}.
This problem of excessive numerical dissipation becomes highly significant in
the case of coupled morphodynamic equations due to the difference in
characteristic time scales. As discussed before, for the coupled system of equations presented
above, the time step is controlled by the more dynamic component of the system i.e.
shallow water equations. This time step is very small for the relatively passive
process of bed evolution and causes excessive smearing of the bed
topography as the solution evolves. In an attempt to remedy this, we present an anti-diffusive,
non-oscillatory (\textit{adNOC}) scheme (see also \citep{zia2015numerical}) that has reduced diffusion and is suitable
for coupled systems such as the morphodynamic equations presented in this study.

Consider the NOC scheme in one dimension:
\[
\overline{u}_{j}^{n+1}=\frac{1}{2}
(\overline{u}_{j+1/2}^{n}+\overline{u}_{j-1/2}^{n})+\frac{\Delta
x}{8}(\sigma_{j-1/2}^{n}-\sigma_{j+1/2}^{n})
\]
\begin{equation}\label{eq:23}
\;\;\;\;-\lambda(f_{j+1/2}^{n+1/2}-f_{j-1/2}^{n+1/2})+\frac{\Delta
t}{2}(s_{j+1/4}^{n+1/2}+s_{j-1/4}^{n+1/2}).
\end{equation}
To demonstrate the anti-diffusive correction, let's assume that the system is at steady state and
there is no source.
This will remove the third and fourth term in Eq. (\ref{eq:23}) leaving just the
first two terms:
\begin{equation}\label{eq:24}
\overline{u}_{j}^{n+1}=\frac{1}{2}
(\overline{u}_{j+1/2}^{n}+\overline{u}_{j-1/2}^{n})+\frac{\Delta
x}{8}(\sigma_{j-1/2}^{n}-\sigma_{j+1/2}^{n}).
\end{equation}
By substituting 
\[
\overline{u}_{j-1/2}^{n}=\frac{1}{2}(\overline{u}_{j-1}^{n-
1}+\overline{u}_{j}^{n-1})+\frac{\Delta
x}{8}(\sigma_{j-1}^{n-1}-\sigma_{j}^{n-1})
\]
and
\[
\overline{u}_{j+1/2}^{n}=\frac{1}{2}(\overline{u}_{j}^{n-
1}+\overline{u}_{j+1}^{n-1})+\frac{\Delta
x}{8}(\sigma_{j}^{n-1}-\sigma_{j+1}^{n-1})
\]
into Eq. (\ref{eq:24}), one obtains
\begin{equation}\label{eq:25}
\overline{u}_{j}^{n+1}=\overline{u}_{j}^{n-1}+\frac{1}{4}
(\overline{u}_{j-1}^{n-1}-2\overline{u}_{j}^{n-
1}+\overline{u}_{j+1}^{n-1})
+\frac{\Delta
x}{8}(\sigma_{j-1}^{n-1}-\sigma_{j+1}^{n-1})+\frac{\Delta
x}{8}(\sigma_{j-1/2}^{n}-\sigma_{j+1/2}^{n}).
\end{equation}
The second term on the right hand side of Eq. (\ref{eq:25}) is the 
finite difference approximation for a diffusive term (ie., $
\frac{1}{4}(\overline{u}_{j-1}^{n-1}-2\overline{u}_{j}^{n-
1}+\overline{u}_{j+1}^{n-1})\approx\frac{(\Delta
x)^{2}}{4}\frac{\partial^{2}u}{\partial x^{2}}$), which shows the origin of
numerical dissipation in this scheme. Notice that the diffusion term is
applied each time a time step is taken. This means that for fixed time period, more time steps
will introduce more numerical dissipation in the solution. This dissipation, however, can be
mitigated or removed entirely by carefully choosing the finite difference approximations of the
slopes present in the equation (see \citep{zia2015numerical} for detail). Substituting
\[
\sigma_{j-1}^{n-1}=\sigma_{j-1/2}^{n}=\frac{\overline{u}_{j}^{n-
1}-\overline{u}_{j-1}^{n-1}}{\Delta x}
\;\;\;and\;\;\;
\sigma_{j+1}^{n-1}=\sigma_{j+1/2}^{n}=\frac{\overline{u}_{j+1}^{n-
1}-\overline{u}_{j}^{n-1}}{\Delta
x}
\]
into Eq. (\ref{eq:25}) results in:
\[
\overline{u}_{j}^{n+1}=\overline{u}_{j}^{n-1}.
\]
Thus, the solution is exactly maintained. The Anti-diffusive, Non-oscillatory
Central difference (\textit{adNOC}) scheme is given by rewriting the NOC scheme with the
anti-diffusive slopes:
\[
\overline{u}_{j+1/2}^{n+1}=\frac{1}{2}
(\hat{u}_{j+1}^{n}+\hat{u}_{j}^{n})+\frac{\Delta x}{8}(1-
\varepsilon)(\sigma_{j}^{n}-\sigma_{j+1}^{n})-\frac{\varepsilon}
{4}(\overline{u}_{j+3/2}^{n-1}-2\overline{u}_{j+1/2}^{n-
1}+\overline{u}_{j-1/2}^{n-1})
\]
\begin{equation}\label{eq:26}
\;\;\;\;-\lambda(f_{j+1}^{n+1/2}-f_{j}^{n+1/2})+\frac{\Delta
t}{2}(s_{j+1/4}^{n+1/2}+s_{j+3/4}^{n+1/2}).
\end{equation}
where $\hat{u}$ is the cell average evaluated without the anti-diffusion
correction i.e. the third term in the right hand side of (\ref{eq:26}).
The new parameter $\varepsilon$ is the factor ($0\leq\varepsilon\leq1$) signifying the strength of the
anti-diffusive slopes. A value of 1 for $\varepsilon$ 
means that only anti-diffusive slopes are used in calculation of reconstructions while a
value of 0 means that the standard NOC scheme is used. Following this
discussion along with the discussion in the previous section, the \textit{adNOC} scheme for the
system (\ref{eq:12}-\ref{eq:16}) is given by:
\[
\overline{W}_{p+1/2,q+1/2}^{n+1}=\frac{1}{4}\{\hat{W}_{p,q}^{n}+\hat{W}_{p+1,q}^{n}+\hat{W}_{p,q+1}^{n}+\hat{W} _{p+1,q+1}^{n}\} \]

\[ +\frac{\Delta x}{16}(1-\varepsilon)\{\sigma_{p,q}^{x}-\sigma_{p+1,q}^{x}- \sigma_{p+1,q+1}^{x}+\sigma_{p,q+1}^{x}\} \]

\[ +\frac{\Delta y}{16}(1-\varepsilon)\{\sigma_{p,q}^{y}+\sigma_{p+1,q}^{y}- \sigma_{p+1,q+1}^{y}-\sigma_{p,q+1}^{y}\}-\varepsilon \Psi \]

\[ -\frac{\Delta t}{2\Delta x}\{F(\overline{W}_{p+1,q}^{n+1/2})+F(\overline{W}_{p+1,q+1}^{n+1/2}) -F(\overline{W}_{p,q}^{n+1/2})-F(\overline{W}_{p,q+1}^{n+1/2})\} \]

\[ -\frac{\Delta t}{2\Delta y}\{G(\overline{W}_{p,q+1}^{n+1/2})+G(\overline{W}_{p+1,q+1}^{n+1/2}) -G(\overline{W}_{p,q}^{n+1/2})-G(\overline{W}_{p+1,q}^{n+1/2})\} \]

\[ +\frac{\Delta t}{4}\{S(\overline{W}_{p+1/4,q+1/4}^{n+1/2})+S(\overline{W}_{p+3/4,q+1/4 }^{n+1/2})+S(\overline{W}_{p+3/4,q+3/4}^{n+1/2})+S(\overline{W}_{p +1/4,q+3/4}^{n+1/2})\}. \]
where $\Psi$ is the 2D anti-diffusive component of the slopes given by:
\[
\Psi=-\frac{3}{4}\overline{W}_{p+1/2,q+1/2}^{n-1}
\]
\[
+\frac{1}{8}
(\overline{W}_{p+1/2,q-1/2}^{n-1}+\overline{W}_{p+1/2,q+3/2}^{n-
1}+\overline{W}_{p-1/2,q+1/2}^{n-1}+\overline{W}_{p+3/2,q+1/2}^{n- 1})
\]
\begin{equation}\label{eq:27}
+\frac{1}{16}(\overline{W}_{p-1/2,q-1/2}^{n-
1}+\overline{W}_{p+3/2,q-1/2}^{n-1}+\overline{W}_{p-1/2,q+3/2}^{n-
1}+\overline{W}_{p+3/2,q+3/2}^{n-1})
\end{equation}
More details on the \textit{adNOC} scheme including several test cases and discussion on
the stability can be seen in \citep{zia2015numerical}. The proof
of the one dimensional anti-diffusive scheme being well-balanced can be seen
in \citep{zia2016}, which can be straightforwardly extended to two dimensions.

\section*{Results}
%\label{sec:Results}

In this section we demonstrate the ability of the \textit{adNOC} scheme to solve the
shallow water equations over a mobile substrate by studying several different
test cases.

\subsection*{Dam Break over erodible bed in one dimension}
%\label{sec:Dam Break 1d}

\begin{figure}
\includegraphics{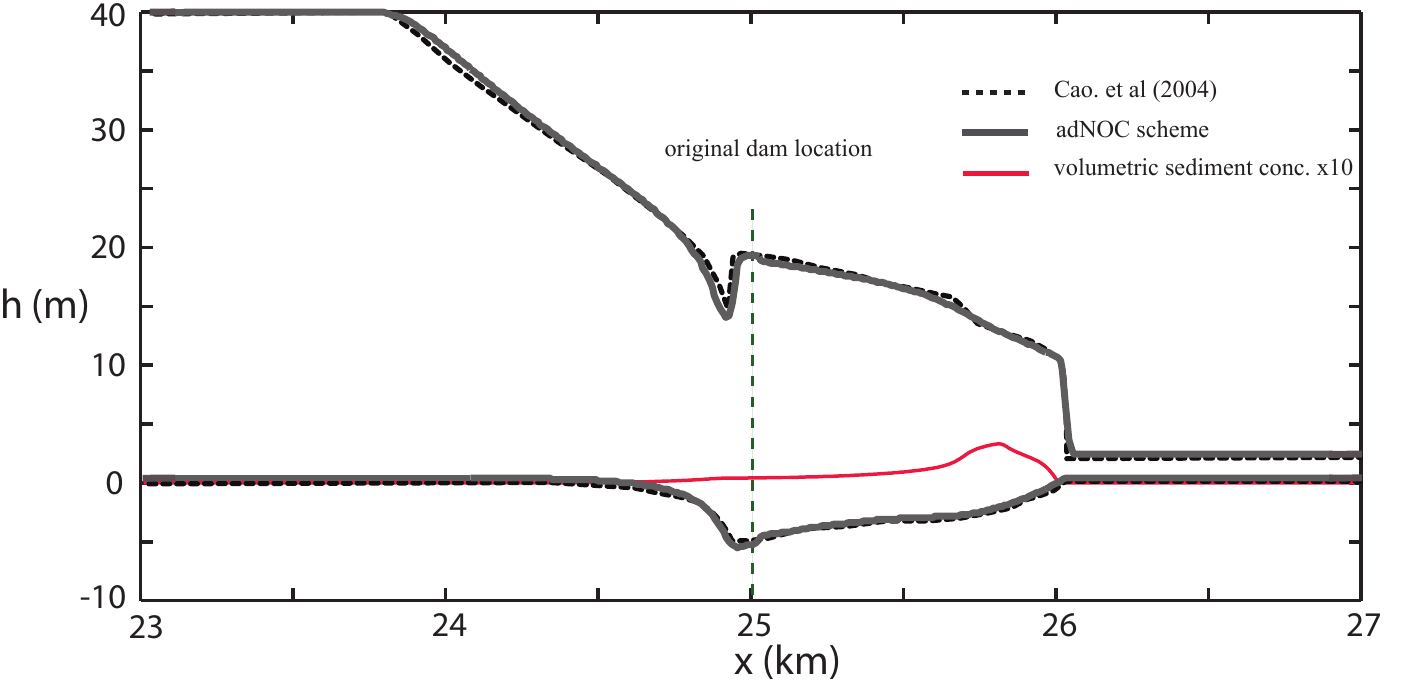}
\caption{1D dam break on erodible bed. Water depth and bed morphology, 60
sec. after dam break.  }
\label{fig:1dcao}
\end{figure}

The performance of the numerical scheme for the solution of the coupled
morphodynamic model is assessed by considering a mobile bed, dam
break problem.
The numerical experiment has been  investigated previously by  \citep{doi:10.1061} and
\citep{Simpson:2006:CMS:1650414.1650589} where the solutions were obtained with upwind methods based on
approximate Riemann solvers. The model setup consists
of a 50 $km$ long, horizontal, one-dimensional channel
with a dam located at 25 $km$ separating two initially stagnant bodies
of water with depths of 40 $m$ and 2 $m$. The dam is breached
instantaneously which results in a sharp shock wave downstream of the initial dam along with
a smooth rarefaction wave upstream within the reservoir. These waves cause the
base to deform in response to erosion and sedimentation. The Values used
for $\varepsilon$ along with other parameters are presented in
Table \ref{table:parametervalues}. We set $\zeta=1$ and  $A=0$, to use the same
empirical formulations as in the previous studies to compare the numerical schemes. Fig.
\ref{fig:1dcao} shows the wave height, bed elevation and sediment concentration near the location of the failed dam after
60 seconds.  Results show the development of a heavily concentrated, eroding
wavefront which forms at the location of the failed dam and diminishes as
it propagates downstream. A hydraulic jump is formed near the previous dam site
due to rapid bed erosion that  reaches a  height of approximately 5 meters. It
can be seen that the sediment transport and bed friction have a strong influence
on the wave height in the dam break scenario. Results presented in Fig.
\ref{fig:1dcao} agree with those computed by \citep{doi:10.1061} and \citep{Simpson:2006:CMS:1650414.1650589} using Riemann-based solvers.

\subsection*{Dam Break over erodible bed in two dimensions}
%\label{sec:Dam Break 2d}

The second test considered is the two dimensional dam break problem introduced by
Cao et al. \citep{cao2010coupled}, which was  designed to reproduce results of a flume experiment.
%The test aims at investigating morphological evolution of an erodable bed made of uniform coarse sand in the dam break scenario.
 The experiment consists of a flume,  3.6 $m$ wide and 36 $m$ long,
 containing an erodible bed made of uniform coarse grained sand.
A one meter wide gate centered across the middle of the
flume is located about 12 $m$ from the upstream end of the flume. The gate
initially separates  a 0.47 $m$ deep upstream water reservoir from an initially
dry bed below the gate.  The bed material consists of 85 $mm$ deep, fully
saturated sand with grain size of 1.61 $mm$ and a  Manning's roughness
coefficient of 0.0165.  A complete list of parameters is provided  in table
\ref{table:parametervalues}.

For the numerical simulation, the boundary conditions are closed wall along the
lateral and  upstream boundaries and a transmissive boundary at the downstream
end. The spatial domain was discretized with 300 cells in the long direction
and 100 cells across the flume. The Courant number was set to 0.4. The bedload
flux was calculated using $A=0.0001$, $b=3$ and the coefficient
$\zeta=0.01$, meaning that sediment transport is dominated by the bedload.

\begin{figure}
	\begin{subfigure}{\textwidth}
		\centering
		
                \includegraphics[scale=0.75]{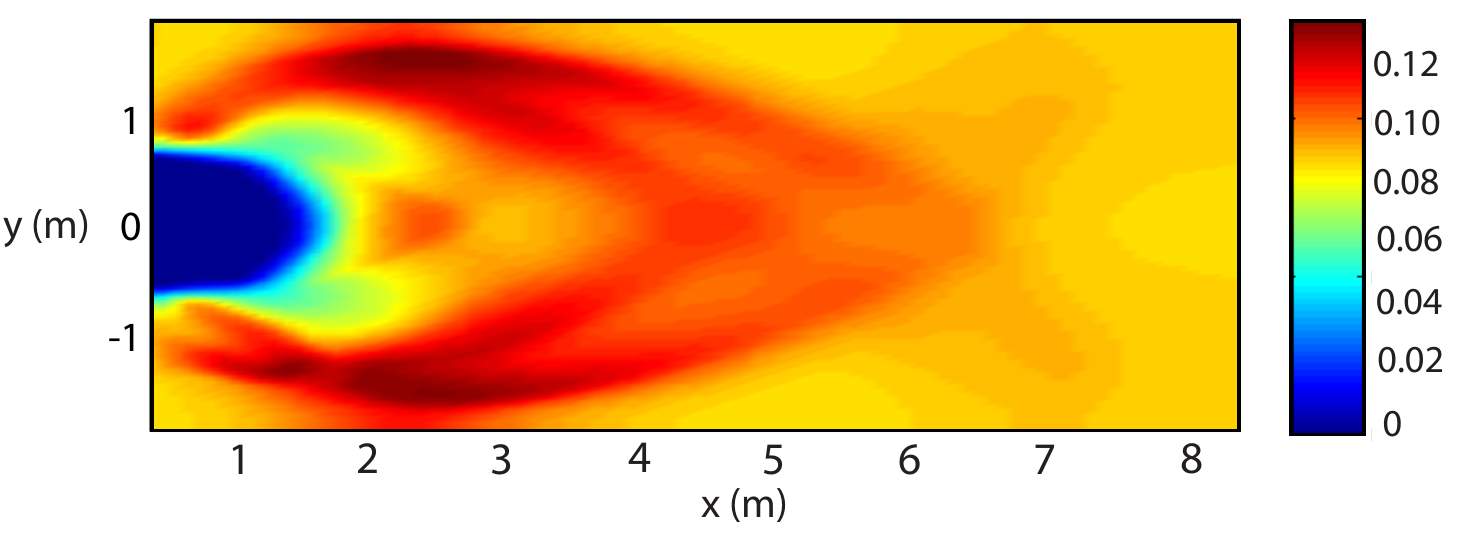}
		\caption{Bed topography 20 sec. after the dam breach  calculated using the
		\textit{adNOC} scheme.}
		\label{fig:cao_adnoc}
	\end{subfigure}
	
	\vspace{15mm}
	\begin{subfigure}{\textwidth}
		\centering
		\includegraphics[scale=0.7]{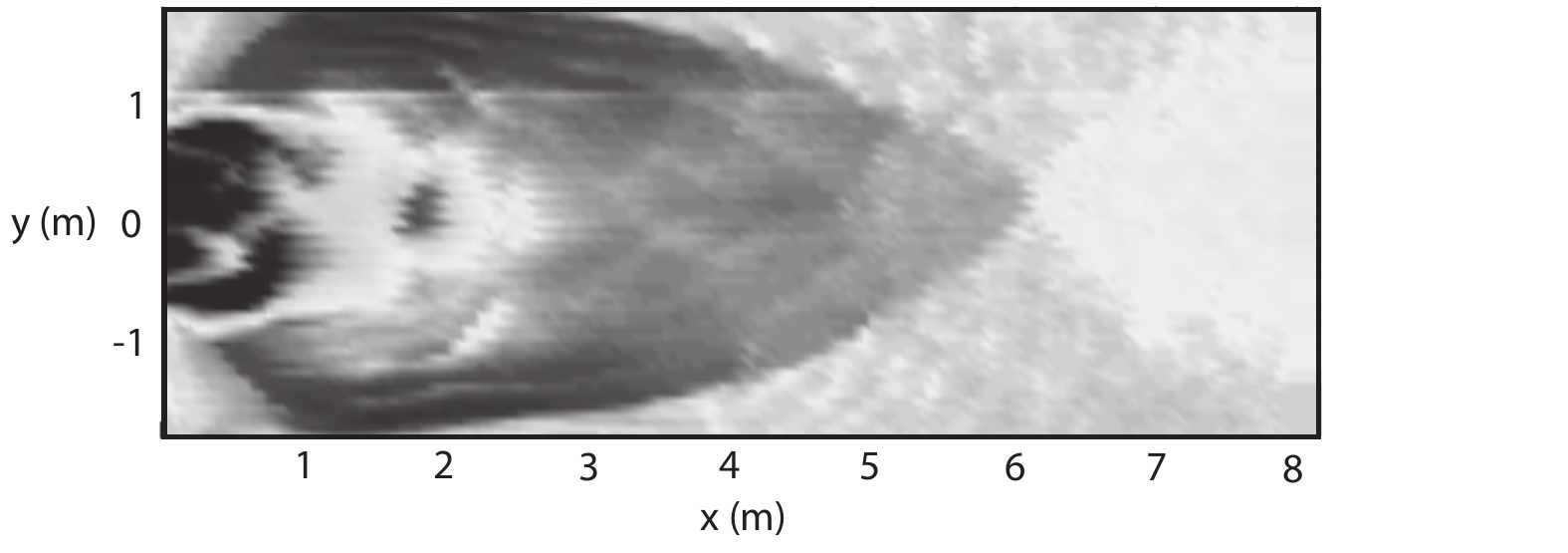}
		\caption{Bed topography 20 sec. after the dam breach determined from a flume
		experiment (published by \citep{cao2010coupled}).}
		\label{fig:cao_exp}
	\end{subfigure}
	
	\vspace{15mm}
	\begin{subfigure}{\textwidth}
		\centering
		\includegraphics[scale=0.85]{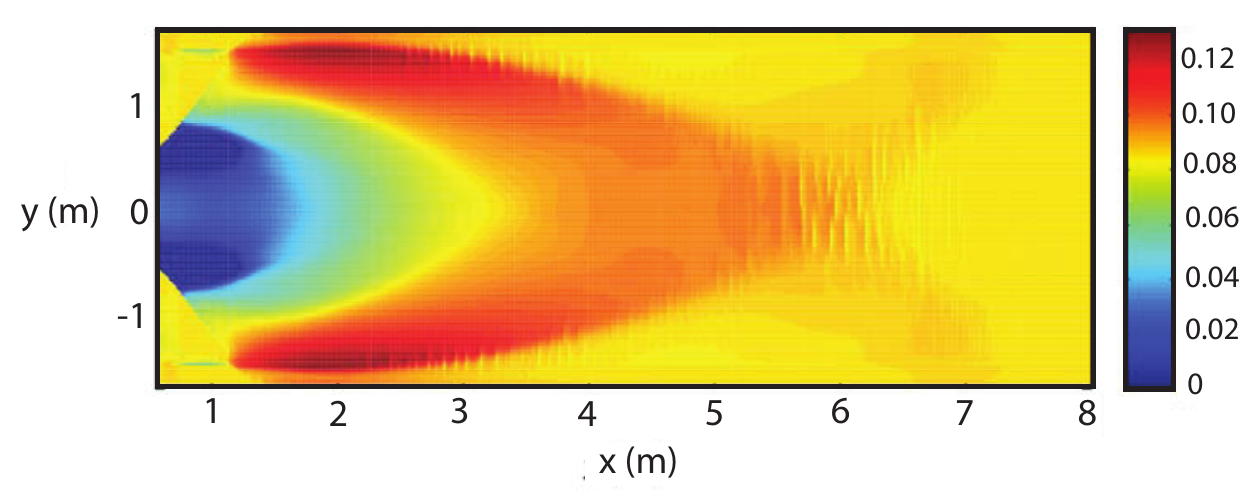}
		\caption{Bed topography 20 sec. after the dam breach as calculated using a
		Riemann-solver based method (published by \citep{cao2010coupled}.}
	\label{fig:cao_num}
	
	\end{subfigure}
	\caption{Two dimensional dam break experiment.}\label{fig:2ddam}
\end{figure}

Fig. \ref{fig:2ddam} shows the  portion of bed topography downstream of the
gate 20 seconds after the breach of the dam computed with the new adNOC scheme 
(\ref{fig:cao_adnoc}), along with a comparison with
the original laboratory experiment (\ref{fig:cao_exp}) and numerical results
computed by \citep{cao2010coupled} with a Riemann based upwind scheme
(\ref{fig:cao_num}). The dam break is seen cause scouring close to the gate,
while a cone shaped dune is deposited further downstream. The  \textit{adNOC}
scheme produces  results consistent with both the laboratory experiment and the Riemann-based numerical solution.

\begin{table}[h]
\small
\begin{tabular}{cccccc}
\hline
\multicolumn{1}{c|}{Parameter}       & dam break 1d       & dam break 2d       & conical dune & sediment wedge & flume experiment   \\ \hline
\textit{g ($m/s^{2}$)}               & 9.8                & 9.8                & 9.8          & 9.8            & 9.8                \\
\textit{i}                           & 2.0                & 2.0                &
-            & -              & 2.0                  \\
$\nu\;(m^{2}/s)$                     & $1.2\times10^{-6}$ & $1.2\times10^{-6}$ & -            & -              & $1.2\times10^{-6}$ \\
$\phi$                               & 0.4                & 0.42               & 0.4          & 0.4            & 0.4                \\
\textit{n}                           & 0.03               & 0.0165             & 0            & 0.082          & 0.033              \\
\textit{s}                           & 1.63               & 1.63               & -            & -              & 1.63               \\
$\theta_{c}$                         & 0.045              & 0.045              & -            & -              & 0.045              \\
\textit{d (mm)}                        & 8                  & 1.61              
& -            & -              & 1                  \\
\textit{f}                           & 0.03               & 0.03               & -            & -              & 0.03               \\
$\varepsilon$ (shallow water system) & 0.6                & 0.6                & 0.3          & 0              & 0                  \\
$\varepsilon$ (suspended transport)  & 0.6                & 0.5                & -            & -              & 0                  \\
$\varepsilon$ (Exner's equation)     & 1                  & 0.9                & 1            & 0.995          & 0.92               \\
$\zeta$                              & 1                  & 0.01               & 0            & 0              & 0.5                \\
\textit{A}                           & 0                  & 0.0001             & 0.001        & 0.04           & 0.001              \\
\textit{b}                           & -                  & 3                  & 3            & 4              & 3                  \\
$\Delta x\;(m)$                      & 10                 & 0.09               & 5            & 0.0225         & 0.5                \\
$\Delta y\;(m)$                      & -                  & 0.036              & 5            & -              & 0.33               \\
\textit{Cn}                          & 0.5                & 0.4                & 0.5          & 0.5            & 0.45              
\end{tabular}
\caption{Parameter values used in the test cases. Units are shown
in brackets where applicable.}\label{table:parametervalues}
\end{table}

\subsection*{Evolution of a conical dune}
%\label{sec:Evolution of a conical dune}

In this test, we investigate the flow of water over a mobile sand dune.  The test
was proposed by Hudson and Sweby \citep{FLD:FLD853} and has been used by many
other studies to assess the performance of two dimensional morphodynamic models (e.g.
\citep{siviglia2013numerical,benkhaldoun2010two,di2009two,Canestrelli2010291}).
The experiment is a good test of the stability and diffusivity of the numerical
scheme as it evaluates very slow evolving bed topography over a long period of
time. The test consists of a 1000$\times$1000 $m$ square domain with an initial
bathymetry given by:
\[
z(x,y,0)=\begin{cases}
0.1+\sin^{2}(\frac{\pi(x-300)}{200})\sin^{2}(\frac{\pi(y-400)}{200}) &
\mbox{if }x\in[300,500],\;y\in[400,600]\\
0.1 & \mbox{if } otherwise
\end{cases}
\]
The boundary conditions are a constant unit discharge of 10 $m^2/s$ on the
inflow upstream and outflow downstream boundaries. The sides are reflective,  free slip boundaries.
 The
initial flow conditions are calculated by applying an initial water level ($h+z$) of 10.1 meters
over the whole of the domain and a boundary flux of 10 $m^2/s$ and 0 $m^2/s$ in
the $x$ and $y$ directions respectively, and by allowing the solution to evolve under fixed base
conditions until a steady state is achieved.
 This steady state is
then taken as time zero for the  movable bed experiment and the deformation of the
sand dune is tracked. We use sediment transport parameters that imply lose
coupling between the bedload and flow. The parameter $A$ in the Grass formula is set to 0.001
and the exponent $b$ is chosen to be 3. In order to compare our results with
other published studies, suspended load is neglected  (i.e., $\zeta=0$). The
parameter $\varepsilon$ is taken as 0.3 for the shallow water system and 1 for
the Exner equation. We use 200 cells in each direction and a  Courant number of
0.5. The values used for all the parameters can be seen in Table \ref{table:parametervalues}.

\begin{figure}
	\centering
	\includegraphics[scale=0.8]{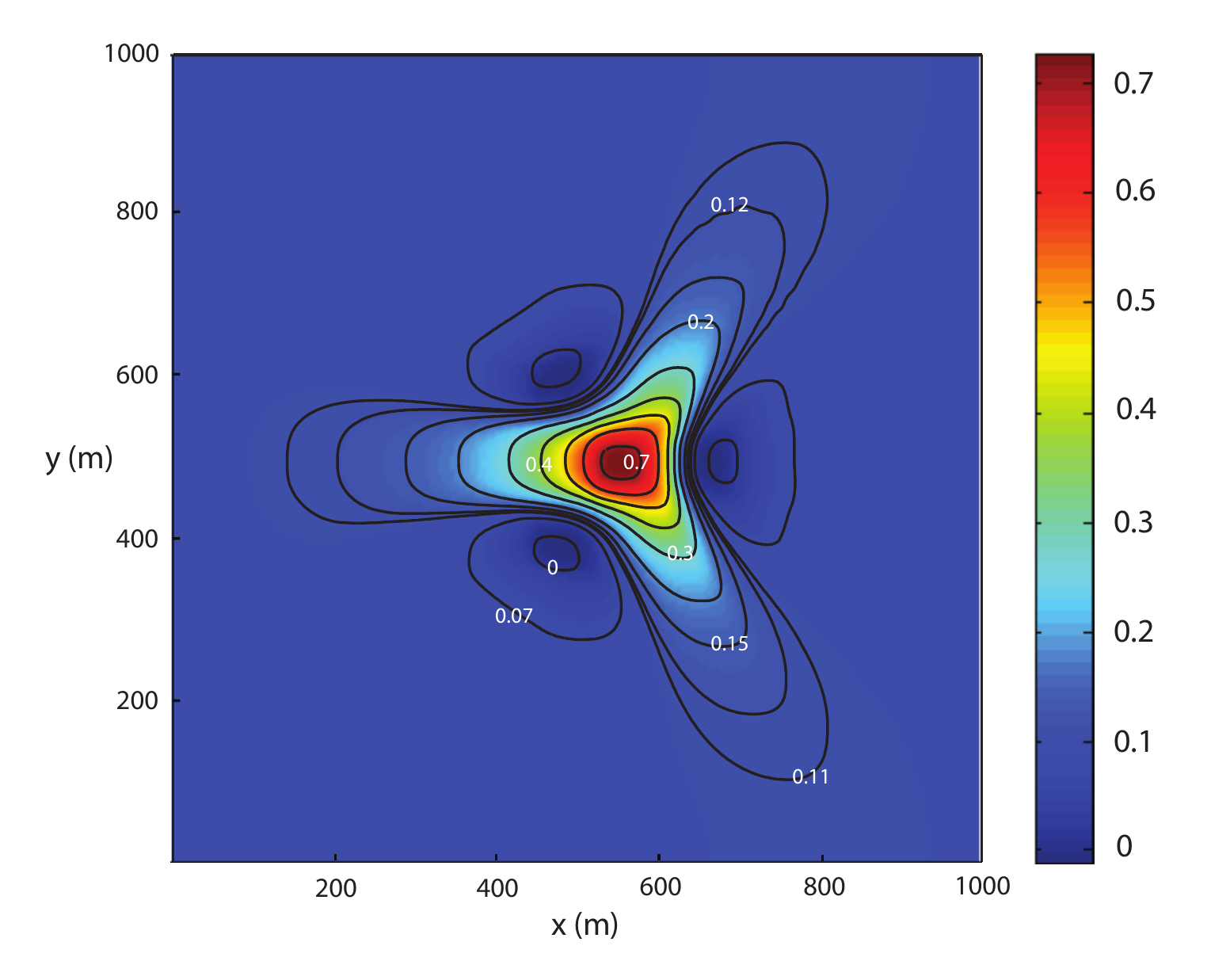}
	\caption{Evolution of a conical dune. Bed topography after
	100 hours.}\label{fig:conical}
\end{figure}

Numerical results, presented  in Fig. \ref{fig:conical}, show that a star shaped
bedform emerges as the result of coupling between sediment transport and the
flowing water. Results computed with the adNOC scheme (\ref{fig:conical})
are similar to the results published by previous studies
(\citep{siviglia2013numerical,benkhaldoun2010two,di2009two}). De Vriend,
in (\citep{de19872dh}), derived an analytical expression to predict the spread angle
of the star shaped pattern under weak flow conditions ($A<0.01$) given by:
\[
\theta=\arctan(\frac{3\sqrt{3}(m-1)}{9m-1})
\]
In this case for $A=0.001$ and $b=3$, the spread angle is predicted  to be
$\theta=21.787^\circ$, which agrees with our numerical results
(24.4$^\circ$).

\subsection*{Stratigraphic development of a one dimensional  sediment wedge}
%\label{sec:Stratigraphic development}

\begin{figure}
	\begin{subfigure}[b]{\textwidth}
		\centering
		\includegraphics[scale=1.25]{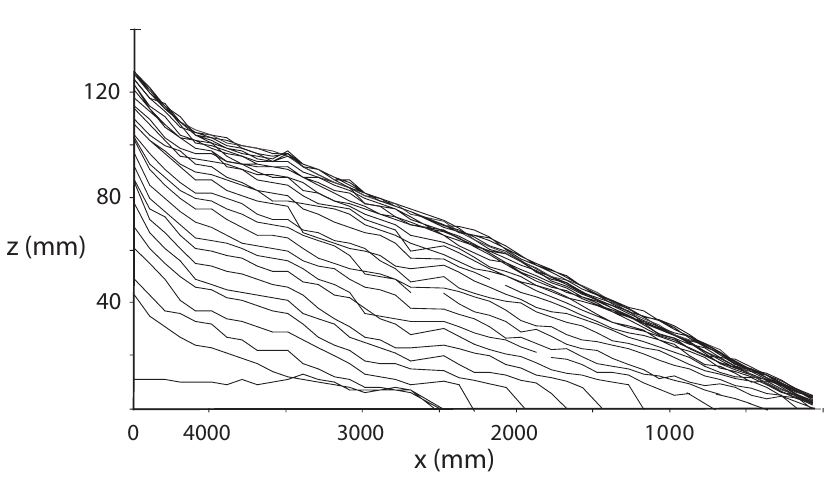}
		\caption{Stratigraphic development of the fluvial sediment wedge in a flume 
		(published by \citep{postma2008sediment}). Each line is drawn after one hour.}
		\label{fig:postma_exp}
	\end{subfigure}

	\begin{subfigure}[b]{\textwidth}
		\centering
		\includegraphics[scale=0.75]{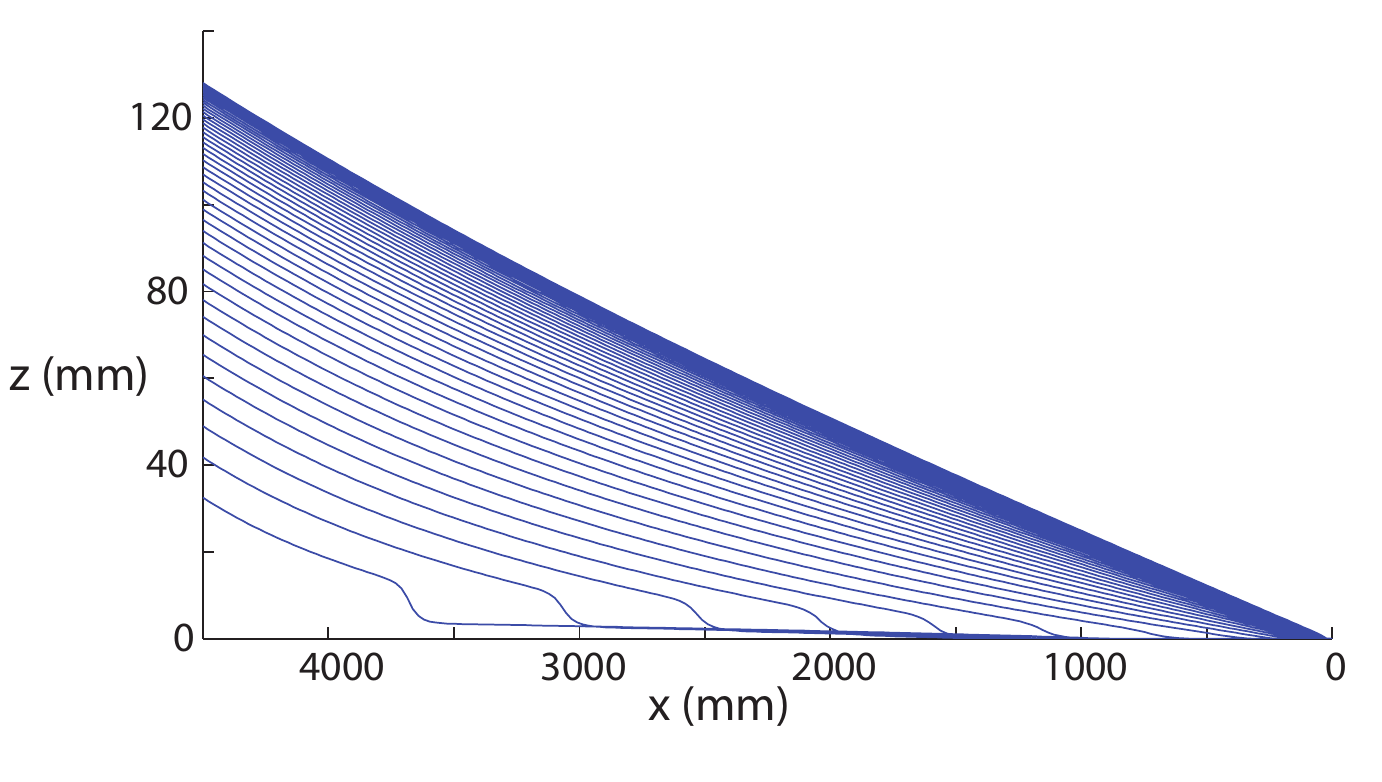}
		\caption{Stratigraphic development as calculated by the morphodynamic model
		using the adNOC scheme. Each line is drawn after one hour.}
		\label{fig:adnoc_postma}
	\end{subfigure}
	\caption{Stratigraphic development in the 1D flume
	experiment.}\label{fig:postma_num}
\end{figure}

 This test is based on a aggradation flume experiment performed by Postma et. al
 \citep{postma2008sediment}. The setup consists of  an initially horizontal 4.5 m long, 0.11 m wide duct,  into which sediment and water are slowly fed.
The water discharge and sediment supply at the  upstream boundary are 350 $dm^3h^{-1}$ and 1.7 $dm^3h^{-1}$, respectively. The sediment has a mean grain size of 250 $\mu m$ and $D_{90}$
of 700 $\mu m$. In the experiments performed by Postma et al. \citep{postma2008sediment}, elevation 
profiles were measured every hour for the first 44 hours (Fig. \ref{fig:postma_exp}).
% Theexperimental measurements published by \citep{postma2008sediment} are shown in Fig. \ref{fig:postma_exp}. %The equilibrium slope is measured to be 0.02857.

We simulate the experiment data using a one dimensional version of the model, which is acceptable because  the width of
the flume is far smaller than its length. Otherwise, the setup and boundary conditions correspond
as closely as possible to those used in the laboratory experiment. The domain is divided into 200 cells and the Courant number is set to  0.5. The
numerical experiment is done with only bedload sediment transport, i.e.
$\zeta=0$ . This makes sense because the water discharge is fairly low and 
suspended transport is  unlikely to occur. The upstream boundary conditions
consist of a fixed water depth (0.01 $m$) and velocity (0.0883 $m/s$). The
downstream boundary is set to be transmissive outflow with the fixed water depth
(0.01 $m$). The parameter values for the Grass bedload model  are $A=0.04$ and
$b=4$ and a Manning roughness coefficient of 0.082 is used for calculation of the friction slope. The rest of the parameter values are shown in Table
\ref{table:parametervalues}. The numerical results presented in Fig.
\ref{fig:postma_num},  show the formation of a sand wedge which propagates progressively downstream with time until the lower boundary is
reached when a steady state bed profile  is achieved. The final equilibrium
slope of the numerical experiments (0.0284) agrees well with that determined
from the flume experiments (i.e., 0.02857) and from a simple analytical model
(0.026).

\subsection*{Two dimensional flume experiment}
%\label{sec:Flume experiment}

The final experiment is done with the aim of demonstrating the
robustness of the \textit{adNOC} scheme for simulating complex flows under near steady
state flow conditions. The experiment consists of a fixed base flume with the
dimensions of 100 $m$ x 10 $m$ and an initial slope of 0.001. Sediment and water
are fed into the flume through a 3 $m$ wide inlet centered along the middle of
the flume at the upstream boundary. Initial water depth is set as 0.2 $m$
throughout the flume. The water at the inlet has a flux of 0.072 $ m^3/s$ and  a
velocity of 0.36 $ m/s$. The lower outflow boundary is set as transmissive with the fixed water depth of 0.2 $m$ while the lateral boundaries are reflective.
The sediment and water mixture fed at the inlet has a volumetric concentration
of 1\% and an additional sediment flux of 0.003 $m^3/s$ is added into the flume
in the form of bedload flux. The sediment is assumed to have a grain size of 1
$mm$ (course sand) and a Manning roughness coefficient of 0.033. The parameters
values used in the experiment are provided in Table \ref{table:parametervalues}.
We have used $\zeta$ as 0.5 and values of $A=0.001$ and $b=3$ are used
to calculate the bedload flux with the Grass formulation. The
numerical calculation presented was performed using 200$\times$30 cells and a
Courant number of 0.45.

\begin{figure}
	\begin{subfigure}[b]{\textwidth}
		\centerline{\includegraphics[scale=0.7]{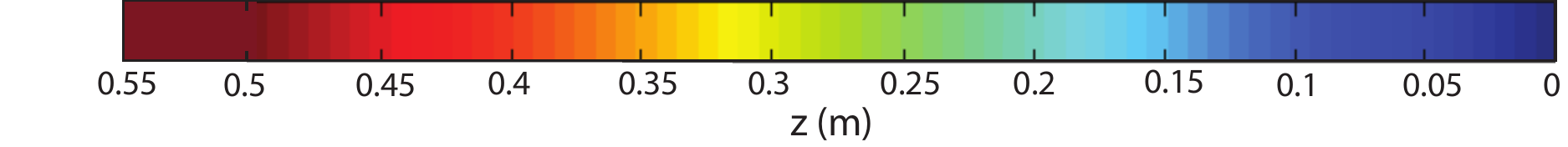}}
	\end{subfigure}
	\begin{subfigure}[b]{\textwidth}
		\centerline{\includegraphics[scale=0.5]{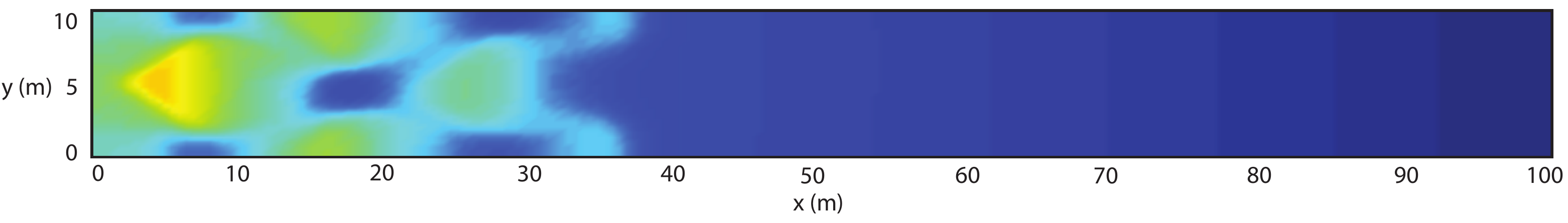}}
		\caption{1.5 hours}
	\end{subfigure}
	\begin{subfigure}[b]{\textwidth}
		\centerline{\includegraphics[scale=0.5]{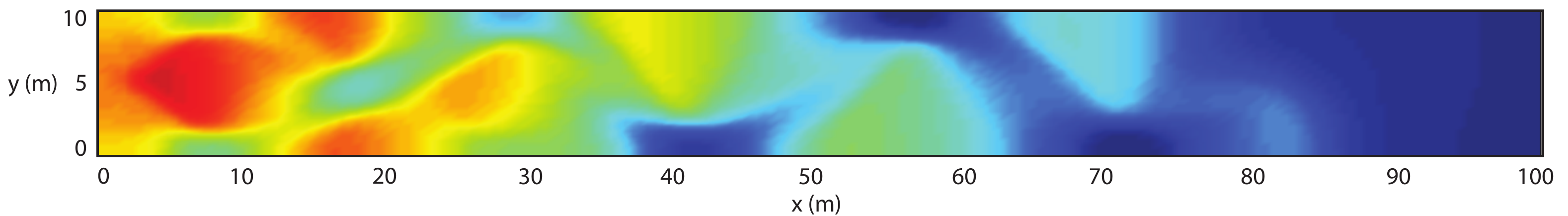}}
		\caption{5 hours}
	\end{subfigure}
	\begin{subfigure}[b]{\textwidth}
		\centerline{\includegraphics[scale=0.5]{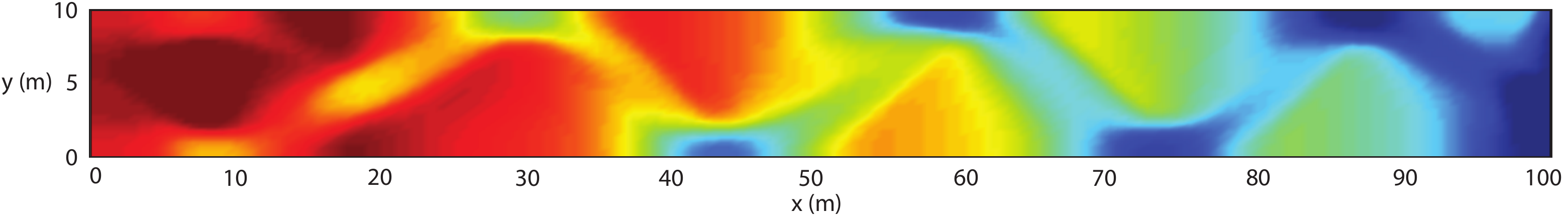}}
		\caption{10 hours}
	\end{subfigure}
	\begin{subfigure}[b]{\textwidth}
		\centerline{\includegraphics[scale=0.5]{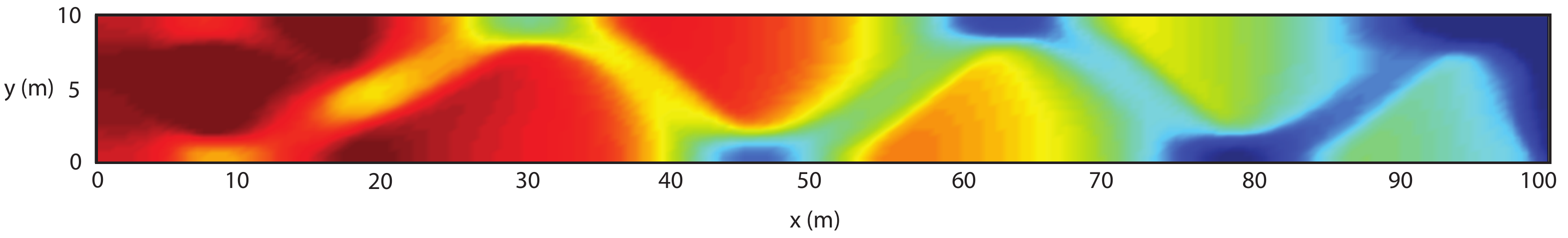}}
		\caption{15 hours}
	\end{subfigure}
	\begin{subfigure}[b]{\textwidth}
		\centerline{\includegraphics[scale=0.5]{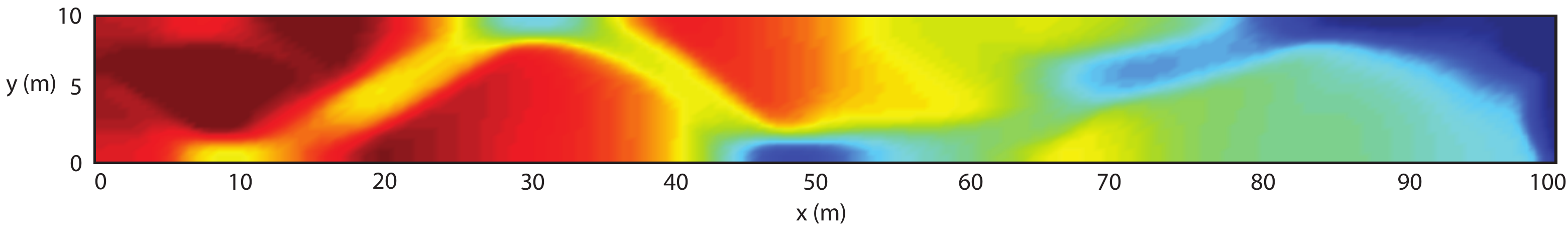}}
		\caption{20 hours}
	\end{subfigure}
	\begin{subfigure}[b]{\textwidth}
		\centerline{\includegraphics[scale=0.5]{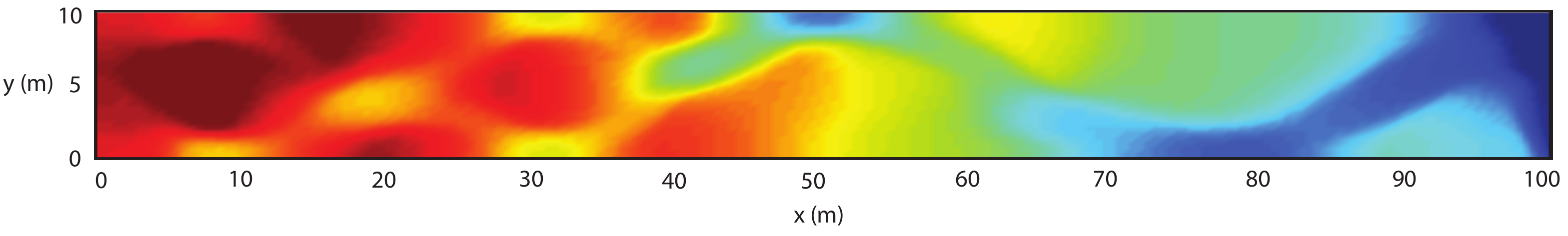}}
		\caption{25 hours}
	\end{subfigure}
	\begin{subfigure}[b]{\textwidth}
		\centerline{\includegraphics[scale=0.5]{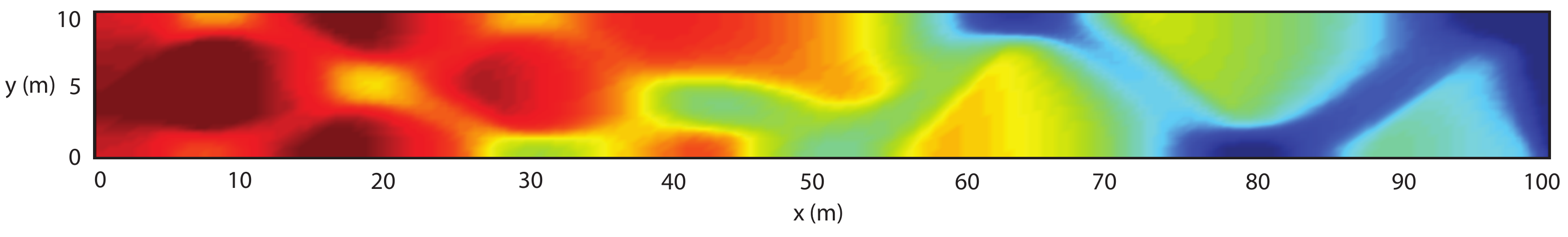}}
		\caption{30 hours}
	\end{subfigure}
\caption{Dynamic bed forms in a two dimensional numerical flume calculated with the adNOC scheme.}\label{fig:flume}
\end{figure}

Fig. \ref{fig:flume} shows the calculated bed topography at
various moments in time.  The results show the formation of complex, dynamic
bed forms due to the strong coupling between the flowing water and sediment
transport. These bedforms are entirely below water. Though it is difficult and
beyond the scope of this work to compare the results directly with either
experimental or natural data, the results show features (e.g., bar formation,
channel avulsion, channel splitting) commonly observed in nature (e.g.
\citep{smith1974sedimentology,field2001channel,kleinhans2013splitting}).
Importantly, the results  illustrate the ability of the \textit{adNOC} scheme
to effectively resolve complex interactions between flow and evolution of the
underlying topography.

\section*{Conclusion} 
\label{sec:Conclusion}

A well-balanced, Anti-diffusive Non-Oscillatory Central Differencing (adNOC) scheme is
presented to solve the shallow water equations coupled to substrate erosion and sediment
transport in two dimensions. The scheme is Riemann solver free and simple to implement. The
complete wave structure of the system is not required and only the largest wave speed 
is used to determine the upper bound on the time step (notably, not to calculate fluxes). This means
that the solution can be evaluated without any knowledge of the wave structure of the sediment transport
system and the empirical relations for erosional/depositional and bedload fluxes can easily be
changed, making it especially suitable for the solution of coupled hydrodynamic/sediment transport
models.

A detailed derivation of the non oscillatory second order scheme is presented demonstrating its
simplicity. The cause of the commonly suffered numerical dissipation by central schemes is
discussed and a correction is proposed. A range of numerical results are presented and 
compared with previously published numerical solutions or with laboratory experiments. The test
cases include highly dynamic one and two dimensional dam break experiments and relatively slow
evolving conical dune and stratigraphic development of a sediment wedge. Simulation of a two
dimensional flume experiment is also performed. Comparison with published results and laboratory
experiments show that the scheme is accurate and robust, suggesting that it has considerable
promise to be used to study a wide range of problems where flow and substrate evolution are coupled.

%\end{linenumbers}
\section*{Notation}
\emph{The following symbols are used in this paper:}
\begin{longtable}{r  @{\hspace{1em}=\hspace{1em}}  l}
$A$ 				& coefficient in Grass formula;\\
$b$ 				& exponent in Grass formula;\\
$c$					& depth-averaged volumetric sediment concentration; \\	
$C_{a}$				& near-bed volumetric sediment concentration;\\
$Cn$				& courant number;\\
$D$					& substrate deposition flux ($m/s$);\\
$d$					& grain diameter ($m$);\\
$E$					& substrate erosion flux ($m/s$);\\
$f$					& Darcy-Weisbach friction factor;\\
$g$					& gravitational acceleration ($m/s^2$);\\
$h$                 & flow depth ($m$);\\
$i$					& exponent in deposition equation;\\
$m$ 				& meter(s);\\
$n$					& Manning's roughness coefficient;\\
$s$					& submerged specific gravity of sediment;\\	
$S_{fx}$			& friction slope in $x$ direction;\\
$S_{fy}$			& friction slope in $y$ direction;\\
$t$                 & time ($s$);\\
$u$					& depth-averaged velocity in $x$ direction ($m/s$);\\
$u_{c}$				& threshold entrainment flow velocity ($m/s$);\\ 	
$u_{\star}$			& friction velocity ($m/s$);\\
$U_{\infty}$		& free surface velocity ($m/s$);\\
$v$					& depth-averaged velocity in $y$ direction ($m/s$);\\
$z$					& bed elevation ($m$);\\
$\alpha_{c}$		& concentration coefficient;\\
$\beta$				& entrainment coefficient;\\
$\varepsilon$ 		& coefficient specifying the strength of anti-diffusive slopes;\\
$\theta$			& Shield's parameter;\\
$\theta_{c}$		& critical value of Shield\textquoteright{}s parameter;\\
$\nu$				& kinematic viscosity of water ($m^2/s$);\\
$\rho$				& density of water-sediment mixture ($kg/m^3$);\\
$\rho_{w}$			& density of water ($kg/m^3$);\\
$\rho_{s}$			& density of sediment ($kg/m^3$);\\
$\rho_{0}$	   		& density of saturated bed ($kg/m^3$);\\
$\sigma^{x}$		& discrete slope in $x$ direction;\\
$\sigma^{y}$		& discrete slope in $y$ direction;\\
$\phi$				& bed sediment porosity;\\
$\omega$			& settling velocity of a single particle in tranquil water ($m/s$);\\
$\zeta$ 			& coefficient in erosion formula;\\

\caption{List of notation used,  with units in brackets where
applicable.}
\label{table:Notations}
\end{longtable}

\pagebreak
\singlespacing
\bibliography{Manuscript}

\end{document}